\documentclass[11pt]{article}

\usepackage[utf8]{inputenc}
\usepackage[T1]{fontenc}
\usepackage{amsmath,amssymb}
\usepackage{graphicx}
\usepackage{geometry}
\usepackage{authblk}
\usepackage{hyperref}
\usepackage{booktabs}
\usepackage{float}
\usepackage{xcolor}
\usepackage[skip=10pt]{caption}
\usepackage[numbers,super,sort&compress]{natbib}

\hypersetup{
  colorlinks=true,
  linkcolor=blue!70!black,
  citecolor=blue!70!black,
  urlcolor=blue!70!black
}

\usepackage{titlesec}
\titleformat{\section}{\normalfont\large\bfseries}{}{0em}{}
\titleformat{\subsection}{\normalfont\normalsize\bfseries}{}{0em}{}
\titleformat{\subsubsection}{\normalfont\normalsize\itshape}{}{0em}{}

\title{\textbf{Physics-Informed 3D Atomic Reconstruction and Dynamics of Free-Standing Graphene from Single Low-Dose TEM Images}}

\author[1,*]{Xiaojun Zhang}
\author[2,*]{Shih-Wei Hung}
\author[1]{Yawei Wu}
\author[3]{Jyh-Pin Chou}
\author[4]{Angus I.\ Kirkland}
\author[5]{Roar Kilaas}
\author[2,*]{Fu-Rong Chen}

\affil[1]{Department of Mechanical Engineering, City University of Hong Kong,
  Kowloon, Hong Kong S.A.R., China}
\affil[2]{Department of Materials Science and Engineering and TRACE EM Unit,
  City University of Hong Kong, Kowloon, Hong Kong S.A.R., China}
\affil[3]{Graduate School of Advanced Technology, National Taiwan
  University, Taipei 10617, Taiwan}
\affil[4]{Department of Materials, University of Oxford, Oxford OX1~3PH, UK}
\affil[5]{Total Resolution LLC, 20 Florida Ave., Berkeley, CA 94707, USA}
\affil[*]{Corresponding authors:
Xiaojun Zhang (\href{mailto:xzhang2365@gmail.com}{xzhang2365@gmail.com});
Shih-Wei Hung (\href{mailto:swhung@cityu.edu.hk}{swhung@cityu.edu.hk});
Fu-Rong Chen (\href{mailto:furong.chen@cityu.edu.hk}{furong.chen@cityu.edu.hk})}

\date{}

\begin{document}
\maketitle

\begin{abstract}

Resolving the three-dimensional (3D) atomic geometry of free-standing graphene in real time is essential for understanding how intrinsic rippling governs its electronic properties. However, the low electron doses required to mitigate radiation damage impose severe signal-to-noise constraints that limit conventional reconstruction methods. Here, we present a physics-informed computational framework that reconstructs 3D atomic coordinates of single-layer graphene from individual low-dose transmission electron microscopy (TEM) frames ($8 \times 10^{3}$\,e$^{-}$/\AA$^{2}$, 1\,ms temporal resolution).  The approach combines simulated annealing optimisation with molecular dynamics regularisation, achieving sub-angstrom out-of-plane accuracy ($\sigma_{z} < 0.45$\,\AA), validated against ground-truth simulations. A Kullback--Leibler divergence-based calibration aligns the forward model with experimental image statistics, reducing systematic bias. Applied to high-speed time-series data, the framework enables simultaneous extraction of real-time ripple dynamics, strain tensors, surface curvature, bond-length distributions, and density functional theory (DFT)-derived electron localisation functions (ELF). We establish quantitative relationships linking local geometry, strain, and bond-length variations to electron localisation, demonstrating that sub-angstrom structural fluctuations drive spatially localised, millisecond-scale electronic modulation. A critical dose threshold is identified below which structural information becomes irrecoverable, providing practical guidance for experimental design. The framework is broadly applicable to beam-sensitive two-dimensional materials.

\end{abstract}

\section{Introduction}

Graphene exhibits exceptional electronic properties, including linear Dirac-cone dispersion, near-ballistic carrier transport, and ultralow resistivity, arising from its $sp^{2}$ bonding network\cite{RN1,RN2,RN3}. These properties are highly sensitive to its three-dimensional (3D) morphology. Free-standing graphene is not strictly planar, but is stabilised by intrinsic out-of-plane rippling with amplitudes below 0.1\,nm and characteristic wavelengths of $\sim$8\,nm\cite{RN4,RN5,RN6}. Such ripples are not merely structural perturbations; they generate local pseudomagnetic fields, modify bond-angle symmetry, and induce spatial variations in the local density of states, ultimately limiting carrier mobility in suspended graphene devices\cite{RN5,RN7,RN8,RN9,RN10,RN11,RN12,RN13}. A quantitative, time-resolved understanding of the full 3D atomic geometry and its direct relationship to electronic structure is therefore essential for controlling graphene-based systems at the atomic scale\cite{RN8,RN9,RN14}.

Aberration-corrected transmission electron microscopy (TEM) enables imaging of individual carbon atoms in graphene with sub-angstrom resolution\cite{RN15,RN16,RN17,RN18,RN19,RN20,RN21,RN22,RN23,RN24}. However, its application to beam-sensitive materials is fundamentally limited by the dual role of the electron beam as both probe and perturbation. At doses sufficient for high signal-to-noise imaging, the beam induces vacancy formation, bond rearrangement, and radiolytic damage\cite{RN23,RN24,RN25,RN26,RN27,RN28,RN29,RN30}. Consequently, imaging must be performed at low electron doses, producing frames with extremely low signal-to-noise ratios (SNR), where conventional 3D reconstruction approaches fail.

This creates a fundamental trade-off between spatial and temporal resolution. Increasing exposure time improves image quality but averages out dynamic processes, while high-speed imaging captures fast dynamics at the cost of further reduced SNR. Existing reconstruction methods, including exit-wave reconstruction\cite{RN16,RN18,RN26}, multi-tilt tomography\cite{RN33,RN34}, and machine-learning-based approaches\cite{RN35,RN36}, typically require multiple frames, higher SNR, or prior structural assumptions that are not available under low-dose, high-speed conditions. 

Here we address this challenge by introducing a physics-informed inverse framework that reconstructs 3D atomic structures directly from single low-dose TEM images. The method combines simulated annealing (SA)\cite{RN37} with molecular dynamics (MD) regularisation to constrain the solution space to physically admissible configurations, enabling robust optimisation under extremely low SNR conditions. In addition, a Kullback--Leibler (KL) divergence-based calibration\cite{RN225} aligns the forward model with experimental image statistics, eliminating systematic bias in the reconstruction process.

Applying this framework to high-speed experimental TEM data, we reconstruct real-time 3D ripple dynamics and establish a comprehensive structure--property analysis pipeline. From the same reconstructed atomic configurations, we extract strain tensors, surface curvature maps, bond-length distributions, and density functional theory (DFT)-derived electron localisation functions (ELF). This enables, for the first time, direct experimental quantification of the relationship between atomic-scale geometry and electronic structure in a dynamically evolving graphene system. Furthermore, we identify a critical electron dose threshold below which structural information becomes irrecoverable, providing practical guidance for low-dose imaging experiments.

\section{Results}

\subsection{Dose calibration and image preprocessing}

Accurate 3D reconstruction requires a forward image model that faithfully reproduces the statistical characteristics of experimental TEM frames. To achieve this, we calibrated the effective electron dose by minimising the Kullback--Leibler (KL) divergence between the experimental reference image and simulated images generated across a range of trial doses.

\begin{figure}[htbp]
  \centering
  \includegraphics[width=\textwidth]{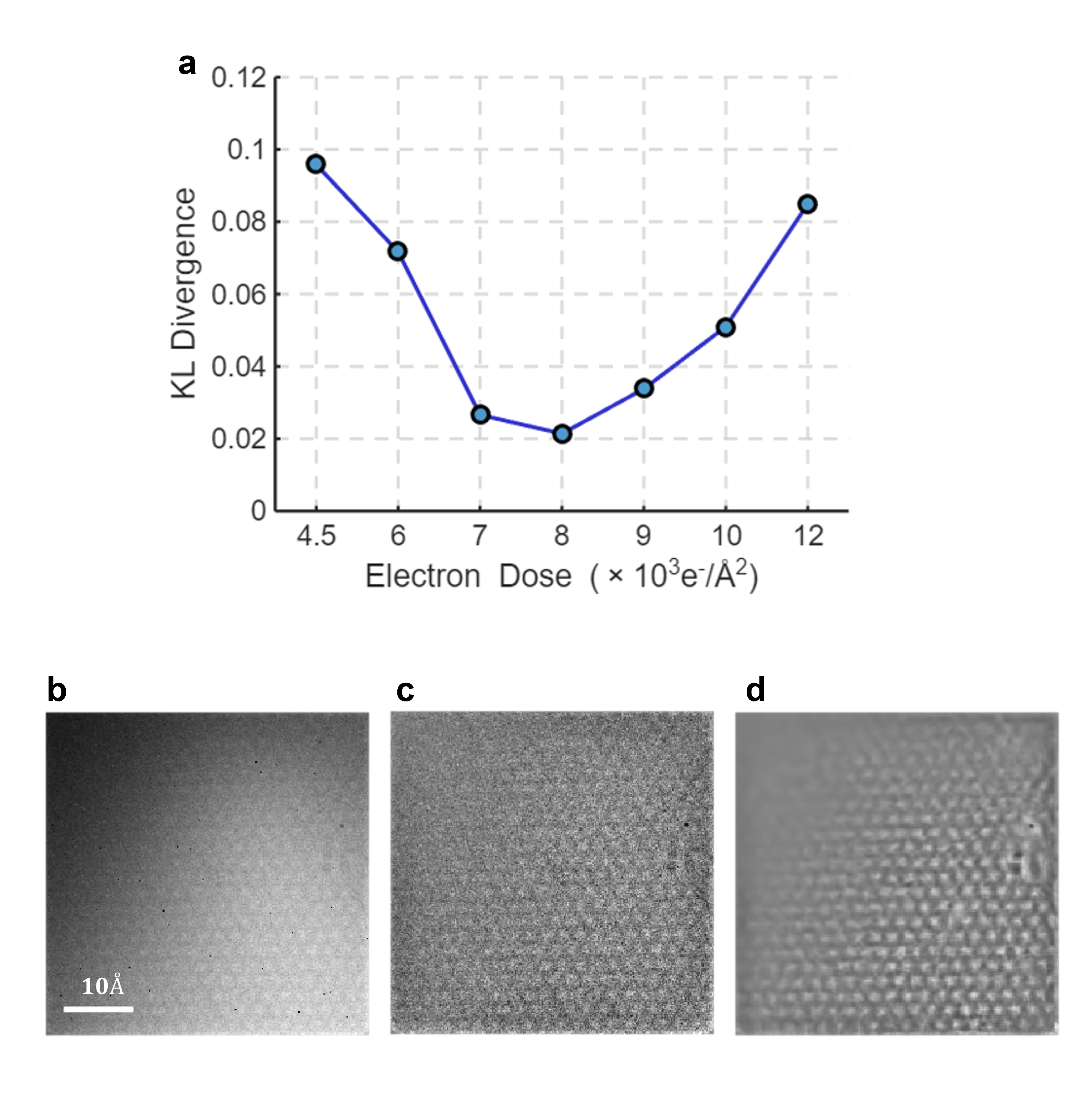}
  \caption{\textbf{Electron dose calibration and image preprocessing.}
  \textbf{a}, KL divergence between the experimental reference frame and
  sparse simulated images at seven trial dose levels; the minimum at
  $8\times10^{3}$\,e$^{-}$/\AA$^{2}$ (highlighted) calibrates the
  forward model (see also Table~\ref{tab:kl}).
  \textbf{b}, Experimental raw TEM frame showing shading distortion and dead pixels($80$\,kV, 1\,ms exposure).
  \textbf{c}, Preprocessed frame after flat-field correction and
  dead-pixel interpolation, ready for structural analysis.
  \textbf{d}, Denoised and temporally averaged over five consecutive frames for the initial structural estimate.}
  \label{fig:1}
\end{figure}

Sparse simulated images were produced at seven dose levels spanning $4.5\times10^{3}$ to $1.2\times10^{4}$\,e$^{-}$/\text{\AA}$^{2}$ (Fig.~\ref{fig:1}a, Table~\ref{tab:kl}). The KL divergence exhibits a clear minimum at $8\times10^{3}$\,e$^{-}$/\text{\AA}$^{2}$ ($D_{\mathrm{KL}} = 0.0214$), which we adopt as the calibrated experimental dose for all subsequent forward simulations.

Raw TEM frames were preprocessed to remove systematic artefacts prior to reconstruction. Flat-field correction was applied using Gaussian background subtraction ($\sigma = 20$ pixels), followed by dead-pixel interpolation using a $5\sigma$ threshold and eight-neighbour averaging. The resulting images were denoised using the BM3D algorithm\cite{RN38} and temporally averaged over five consecutive frames to stabilise the initial structural estimate (Fig.~\ref{fig:1}b--d).

\begin{table}[htbp]
  \centering
  \caption{\textbf{KL divergence between the experimental reference frame
  and sparse simulated images at seven trial doses.}  The minimum at
  $8\times10^{3}$\,e$^{-}$/\AA$^{2}$ identifies the experimental
  operating dose used in all forward simulations.}
  \label{tab:kl}
  \begin{tabular}{lrrrrrrr}
    \toprule
    Dose ($\times10^{3}$\,e$^{-}$/\AA$^{2}$)
      & 4.5 & 6.0 & 7.0 & \textbf{8.0} & 9.0 & 10.0 & 12.0 \\
    \midrule
    $D_{\mathrm{KL}}$
      & 0.096 & 0.072 & 0.027 & \textbf{0.021} & 0.034 & 0.051 & 0.085 \\
    \bottomrule
  \end{tabular}
\end{table}

\subsection{Validation of reconstruction accuracy on simulated data}

The reconstruction pipeline estimates in-plane atomic coordinates ($x$, $y$) by using multiple Gaussian fitting and initialises out-of-plane coordinates ($z$) using the projected charge density (PCD) approximation\cite{RN23}, followed by LOWESS smoothing\cite{RN59,RN60}. This initial model is subsequently refined through simulated annealing (SA), which minimises the pixel-wise $\chi^{2}$ discrepancy between forward-simulated and experimental TEM images. To ensure physically realistic configurations under low signal-to-noise conditions, molecular dynamics (MD) relaxation (LAMMPS, Tersoff potential\cite{RN45,RN46,RN47}) is applied after each SA update, acting as a physics-based regularisation step (see Methods).

\begin{figure}[htbp]
  \centering
  \includegraphics[width=0.95\textwidth]{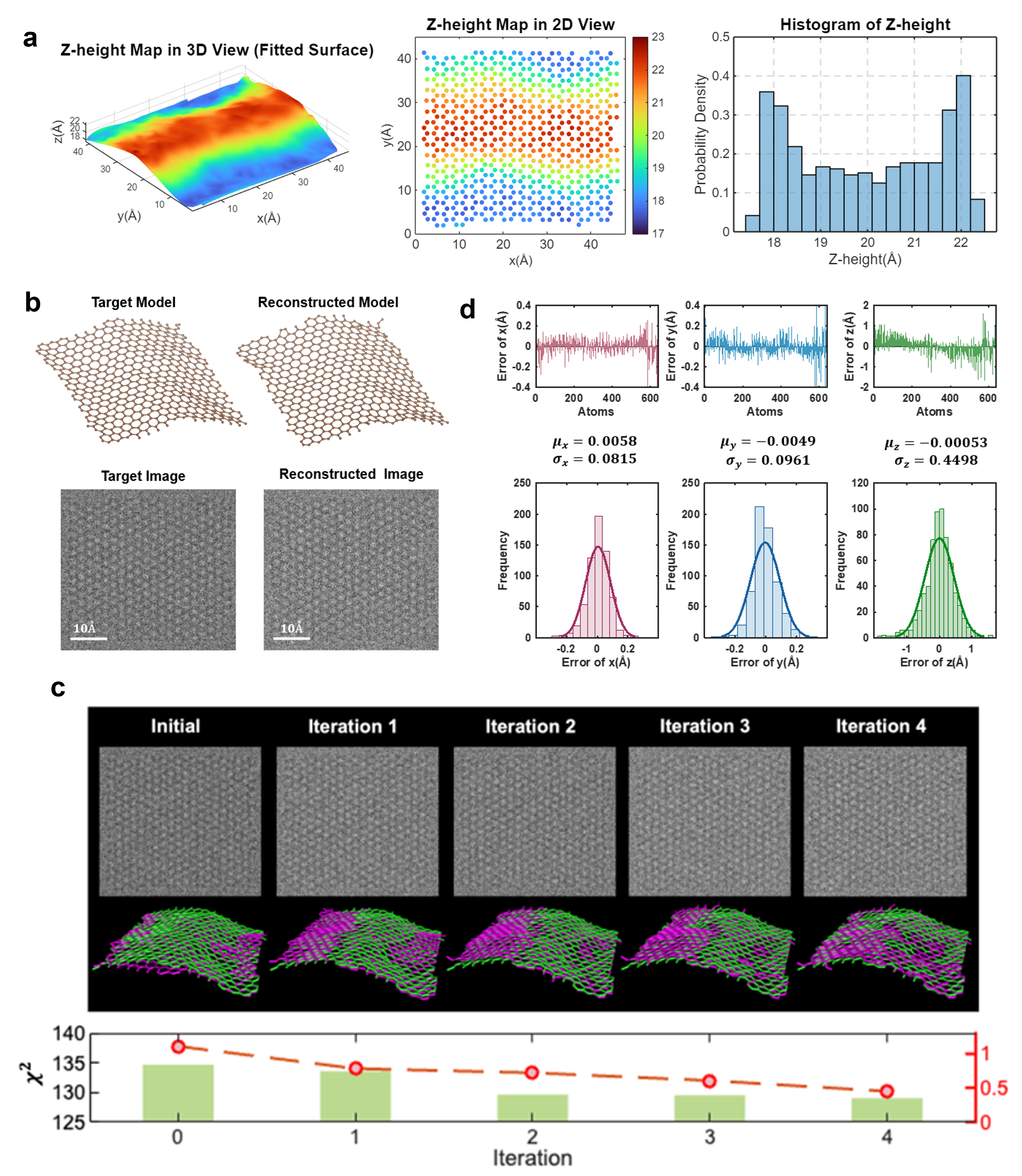}
  \caption{\textbf{Validation of 3D reconstruction accuracy on simulated
  data.}
  \textbf{a}, Ground-truth 640-atom graphene model from MD simulation
  shown from 3D and 2D perspectives with $z$-height colormap and
  histogram.
  \textbf{b}, Ground-truth model and synthetic target TEM image at
  $8\times10^{3}$\,e$^{-}$/\AA$^{2}$ (left); final reconstructed model
  and its forward-simulated image (right).
  \textbf{c}, $\chi^{2}$ (green bars) and $z$-RMSD (red line) vs.\
  iteration number; insets compare the reconstructed model (purple) with
  ground truth (green) at each step alongside the simulated TEM images.
  \textbf{d}, Per-atom reconstruction errors in $x$, $y$, $z$ with
  Gaussian fits; $x$--$z$ projection demonstrates convergence of the
  ripple profile.  In-plane errors are $<$0.1\,\AA;
  $\sigma_{z} = 0.45$\,\AA.}
  \label{fig:2}
\end{figure}

We first validate the method using simulated TEM data generated from a 640-atom graphene model with known ground truth. The reconstruction converges rapidly within four SA iterations, reducing $\chi^{2}$ from 134.7 to 129.0 and the out-of-plane root mean square deviation (RMSD) from 1.11\,\AA\ to 0.45\,\AA\ (Fig.~\ref{fig:2}, Table~\ref{tab:validation}).

In-plane reconstruction errors remain significantly smaller ($\sigma_{x} = 0.082$\,\AA, $\sigma_{y} = 0.096$\,\AA), consistent with the fact that the $z$ direction is underdetermined in single-projection imaging. The achieved out-of-plane accuracy of 0.45\,\AA\ exceeds the performance of previously reported single-image reconstruction approaches at comparable electron dose levels.

These results demonstrate that the integration of SA optimisation with MD-based physical constraints enables stable and accurate reconstruction under realistic low-dose imaging conditions.

\begin{table}[htbp]
  \centering
  \caption{\textbf{Convergence of SA reconstruction on simulated data.}
  $\chi^{2}$ and $z$-RMSD at each iteration; the final accuracy of
  $0.45$\,\AA\ represents a 2.5$\times$ improvement over the
  initialisation.}
  \label{tab:validation}
  \begin{tabular}{lrrrrr}
    \toprule
     & Initial & Iter.\ 1 & Iter.\ 2 & Iter.\ 3 & Iter.\ 4 \\
    \midrule
    $\chi^{2}$       & 134.7 & 133.6 & 129.6 & 129.5 & 129.0 \\
    $z$-RMSD (\AA)   & 1.11  & 0.78  & 0.72  & 0.61  & 0.45  \\
    \bottomrule
  \end{tabular}
\end{table}

\subsection{Real-time 3D ripple dynamics}

We next apply the framework to experimental high-speed TEM data (80\,kV, 1\,ms per frame, $\approx 8\times10^{3}$\,e$^{-}$/\text{\AA}$^{2}$), reconstructing 3D atomic coordinates independently for five consecutive frames from a region containing $\sim$747 carbon atoms.

\begin{figure}[htbp]
  \centering
  \includegraphics[width=\textwidth]{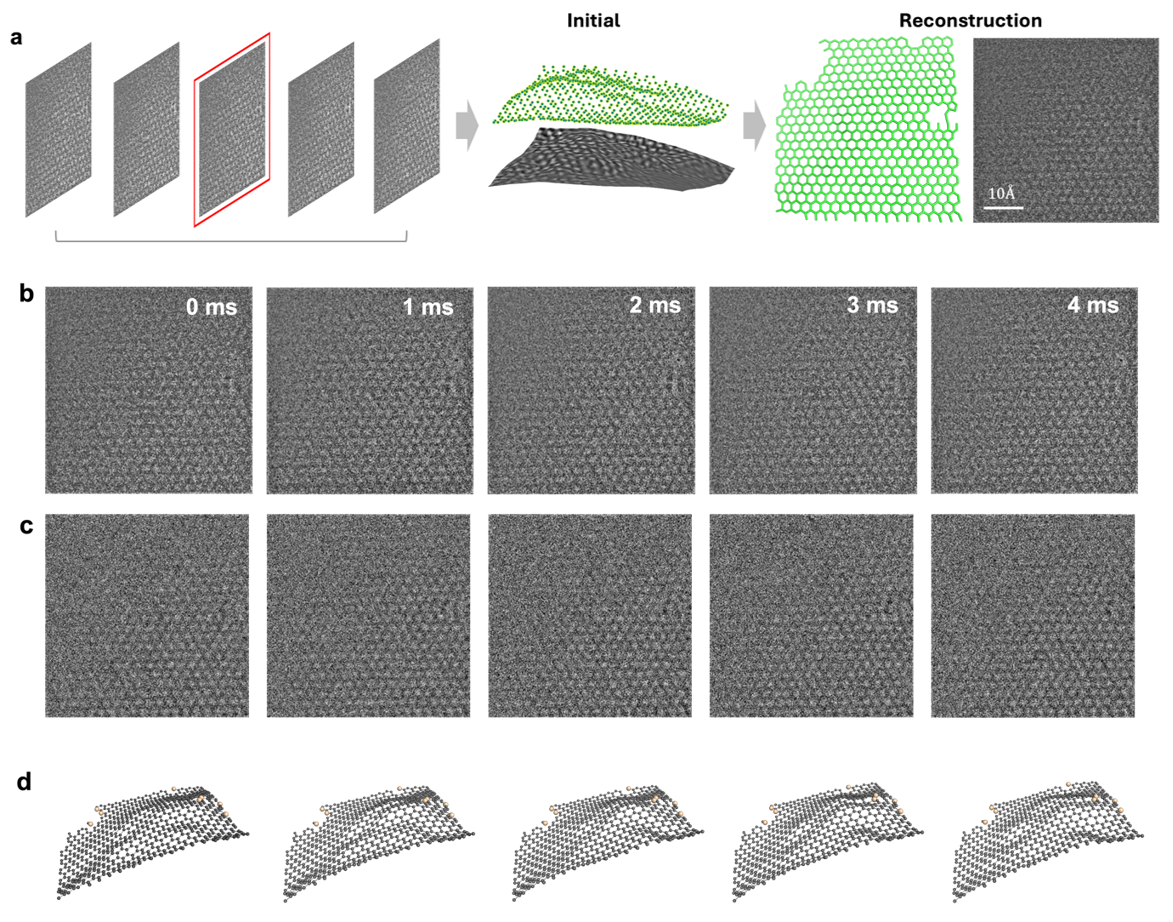}
  \caption{\textbf{Real-time 3D reconstruction of graphene ripple
  dynamics.}
  \textbf{a}, Preprocessing pipeline: the target frame (red rectangle)
  is averaged over five consecutive denoised experimental images to
  produce a stable initial model (centre); the right panel shows the
  final reconstructed atomic arrangement and its forward-simulated TEM
  image.
  \textbf{b}, Five consecutive experimental HRTEM frames (0--4\,ms),
  each containing $\approx$747 carbon atoms at
  $\approx8\times10^{3}$\,e$^{-}$/\AA$^{2}$ per frame.
  \textbf{c}, TEM images simulated from each reconstructed 3D model
  (Tempas), demonstrating close agreement with experimental contrast.
  \textbf{d}, Reconstructed 3D atomic models at each time step,
  revealing millisecond-scale evolution of out-of-plane ripples}
  \label{fig:3}
\end{figure}

The reconstructed structures reveal pronounced out-of-plane displacements of $\Delta z = \pm 4$\,\AA\ relative to the fitted central surface, with ripple morphology evolving significantly on the millisecond timescale under electron-beam excitation (Fig.~\ref{fig:3}). Forward-simulated TEM images generated from the reconstructed 3D models closely reproduce the experimental contrast (Fig.~\ref{fig:3}c), confirming the structural fidelity of the reconstruction. The achieved temporal resolution of 1\,ms enables direct observation of dynamic ripple evolution that would otherwise be averaged out in conventional longer-exposure imaging. These results demonstrate that the proposed framework can capture real-time 3D structural dynamics in beam-sensitive materials under low-dose conditions.

For subsequent analysis, two complementary representations of the reconstructed structures are defined. The \textit{non-flattened model} preserves the full 3D morphology and is used for curvature, bond-length, and electronic analyses. The \textit{flattened model}, obtained by removing global sample tilt through least-squares plane fitting, enables intrinsic in-plane strain analysis by isolating local lattice distortions from global geometric effects.

\subsection{Strain tensor analysis}

To isolate intrinsic lattice deformation from global geometric effects, we analyse the reconstructed structures in both non-flattened and flattened representations.

\begin{figure}[htbp]
  \centering
  \includegraphics[width=0.9\textwidth]{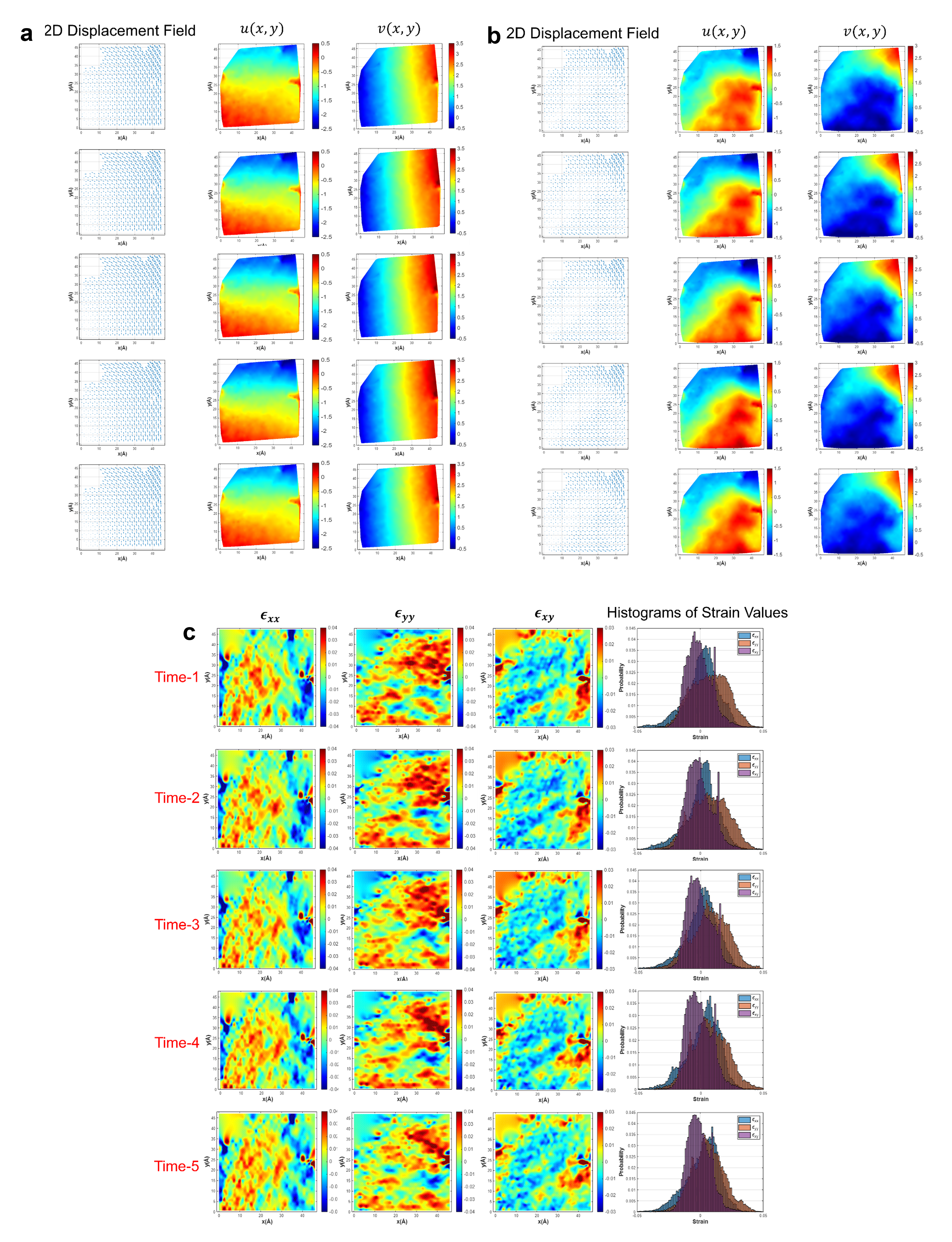}
  \caption{\textbf{Strain tensor analysis at five time steps (0--4\,ms).}
  \textbf{a}, Displacement maps (quiver, $x$-component, $y$-component)
  for the non-flattened (tilted) models: uniform gradients dominated by
  global tilt obscure intrinsic deformation.
  \textbf{b}, Displacement maps for tilt-corrected (flattened) models:
  displacements reduce to $\pm0.15$\,\AA, exposing localised lattice
  distortions.
  \textbf{c}, Full strain maps ($\epsilon_{xx}$, $\epsilon_{yy}$,
  $\epsilon_{xy}$, per-frame histogram) for flattened models.  Shear
  strain $\epsilon_{xy}$ reaches $\pm0.04$ at high-curvature ripple
  flanks and evolves on the millisecond timescale.}
  \label{fig:4}
\end{figure}

In the non-flattened model, displacement fields are dominated by global sample tilt, resulting in nearly uniform strain components within the range $(-0.01, +0.01)$ (Fig.~\ref{fig:4}a). This masks local structural variations and limits quantitative interpretation. After tilt correction via least-squares plane fitting, the flattened model reveals intrinsic lattice distortions. Atomic displacements are reduced from $\pm 0.5$\,\AA\ to $\pm 0.15$\,\AA, and spatially localised strain features emerge (Fig.~\ref{fig:4}b--c). The most prominent feature is the shear strain component $\epsilon_{xy}$, which reaches values up to $\pm 0.04$ at regions of rapid out-of-plane variation. These regions coincide with the flanks of surface ripples identified in subsequent gradient analysis. In contrast, the normal strain components $\epsilon_{xx}$ and $\epsilon_{yy}$ remain comparatively small, indicating that shear deformation is the dominant mode of lattice distortion in rippled free-standing graphene.

The spatial distribution of strain evolves dynamically across consecutive frames, reflecting the coupling between out-of-plane morphology and in-plane lattice deformation under electron-beam excitation. These results demonstrate that ripple-induced curvature directly drives local shear strain at the atomic scale.

\subsection{Surface curvature, bond-length mapping, and electron localisation}

We next quantify the coupling between local geometry and electronic structure by extracting three key descriptors from the reconstructed atomic configurations: surface curvature, bond-length variation, and electron localisation.

The local surface morphology is characterised by the gradient magnitude 
\[
g = \sqrt{\left(\frac{\partial F}{\partial x}\right)^2 + \left(\frac{\partial F}{\partial y}\right)^2},
\]
where $F(x,y)$ is the interpolated height field. This quantity captures the local slope of the graphene sheet and identifies regions of strong curvature.

Bond-length variations are described by the relative deviation
\[
\delta = \frac{b_i - b_0}{b_0},
\]
where $b_0 = 1.42$\,\AA\ is the equilibrium C--C bond length. Spatial maps of $\delta$ reveal regions of local lattice stretching and compression.

\begin{figure}[htbp]
  \centering
  \includegraphics[width=\textwidth]{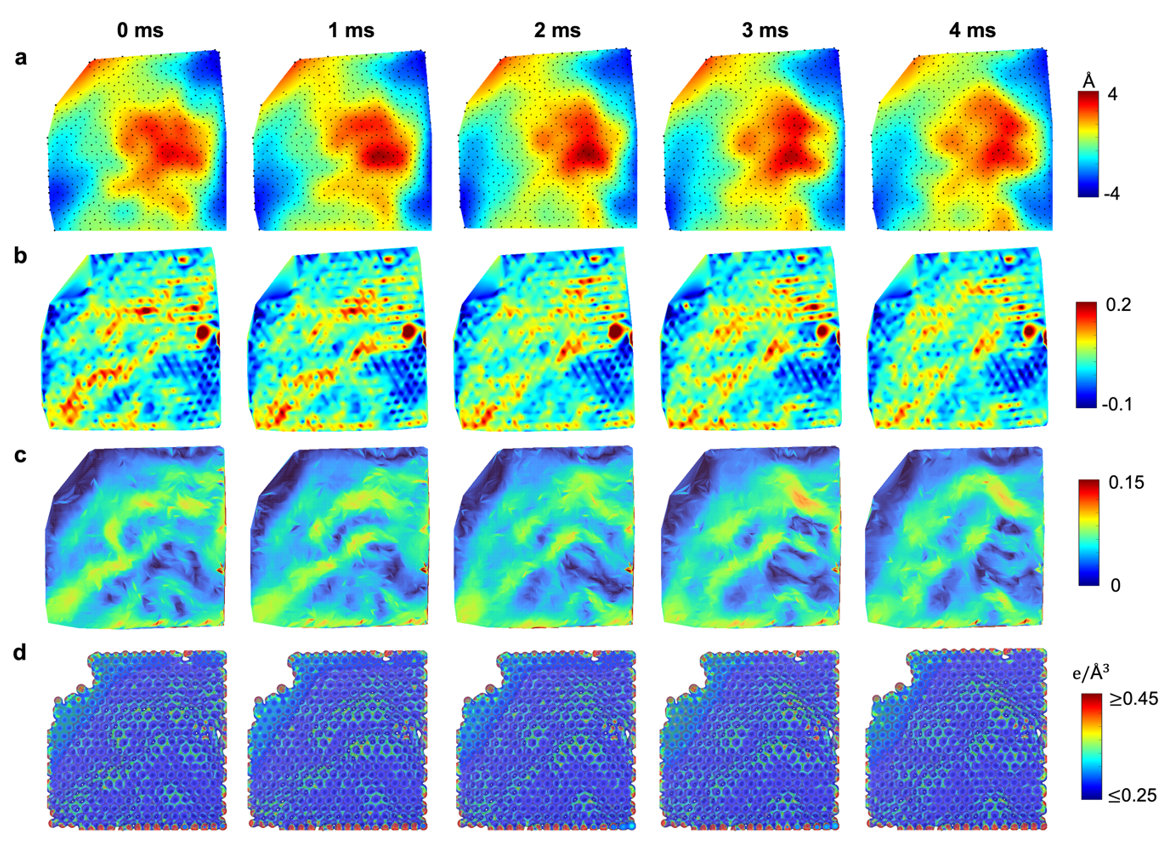}
  \caption{\textbf{Coupled geometric and electronic characterisation at
  five time steps (0--4\,ms).}
  \textbf{a}, 2D maps of out-of-plane displacement $\Delta z$ relative
  to fitted central surface $f_{0}$; values span $-4$ to $+4$\,\AA.
  \textbf{b}, Bond-length change maps $\delta=(b_{i}-b_{0})/b_{0}$
  ($b_{0}=1.42$\,\AA) with 2D spatial distribution and per-frame
  histograms; bond elongation concentrates at high-curvature zones.
  \textbf{c}, Surface gradient magnitude $g$, identifying ripple flanks
  and regions of strong local curvature that evolve between frames.
  \textbf{d}, ELF distributions from DFT, with higher values
  ($\geq$0.45) at high-curvature, elongated-bond regions, indicating
  reduced $\pi$-electron delocalisation.}
  \label{fig:5}
\end{figure}

To probe the electronic response to these structural distortions, density functional theory (DFT) calculations were performed for each reconstructed configuration to obtain the electron localisation function (ELF)\cite{RN56,RN57}. The ELF provides a measure of $\pi$-electron localisation, with higher values indicating reduced orbital overlap and increased localisation. The combined analysis (Fig.~\ref{fig:5}) reveals a clear spatial correlation between geometric distortion and electronic structure. Regions with high surface gradient --- corresponding to ripple flanks --- consistently exhibit increased bond-length variation and elevated ELF values. This indicates that local curvature reduces $p$-orbital overlap, leading to enhanced electron localisation. Importantly, this spatial pattern evolves dynamically across consecutive frames, demonstrating that millisecond-scale morphological fluctuations induce transient, spatially localised modulation of electronic properties. The strong correspondence between $\Delta z$, $\delta$, $g$, and ELF provides direct experimental evidence of structure--property coupling in dynamically rippling graphene.

\subsection{Quantitative structure--electronic property relationships}

\begin{figure}[htbp]
  \centering
  \includegraphics[width=0.95\textwidth]{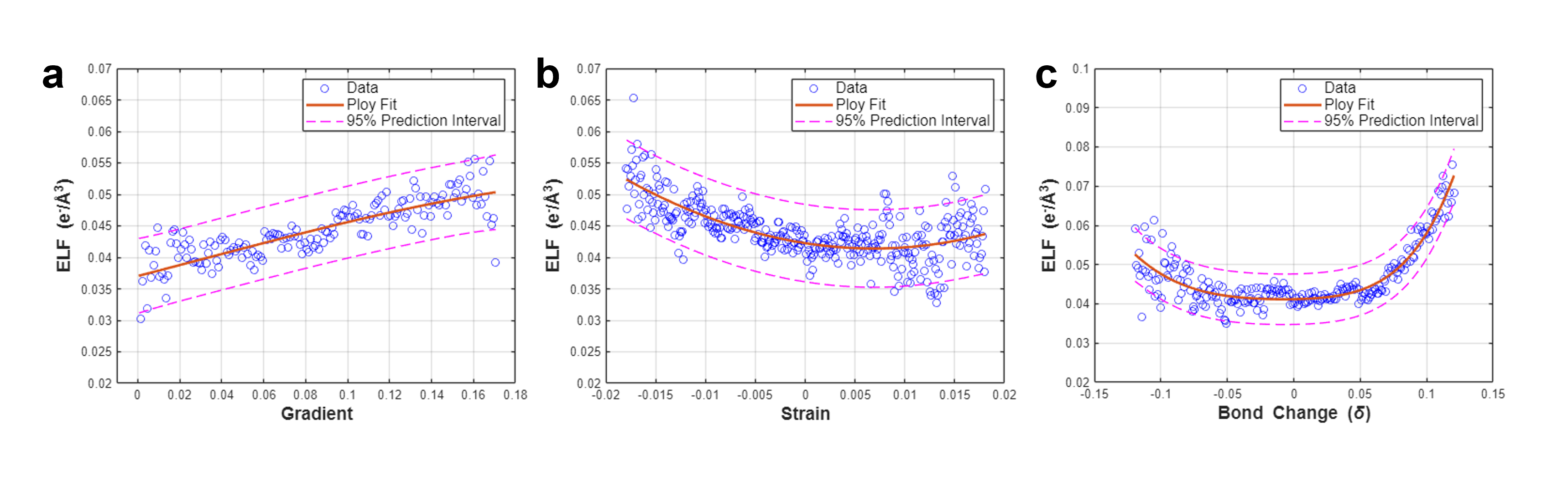}
  \caption{\textbf{Quantitative polynomial relationships between local
  geometry and electron localisation.}
  \textbf{a}, ELF vs.\ surface gradient magnitude $g$
  (Eq.~\ref{eq:elf_g}): non-monotonic dependence reflecting competition
  between curvature-induced bond elongation and compressive effects at
  ripple crests.
  \textbf{b}, ELF vs.\ shear strain $\epsilon_{xy}$
  (Eq.~\ref{eq:elf_e}): symmetric cubic dependence from $sp^{2}$
  bond-angle symmetry breaking.
  \textbf{c}, ELF vs.\ bond-length change $\delta$
  (Eq.~\ref{eq:elf_d}): sharp threshold at $\delta\approx0.1$, marking
  the onset of significant $\pi$-electron localisation.
  In all panels: red curve = polynomial fit; purple dotted lines =
  95\% confidence bounds.  These relationships are the first quantitative
  structure--ELF mappings derived from experimentally reconstructed,
  dynamically evolving atomic configurations.}
  \label{fig:6}
\end{figure}

We quantified the dependence of ELF on each geometric variable
independently by polynomial regression on one representative
reconstructed frame (Fig.~\ref{fig:6}).  Three relationships emerge,
each with distinct physical content:

\begin{equation}
  \mathrm{ELF}(g) = -0.58\,g^{3} + 0.05\,g^{2} + 0.09\,g + 0.04
  \label{eq:elf_g}
\end{equation}
\begin{equation}
  \mathrm{ELF}(\epsilon_{xy}) = 14.64\,\epsilon_{xy}^{3}
    + 17.91\,\epsilon_{xy}^{2} - 0.25\,\epsilon_{xy} + 0.04
  \label{eq:elf_e}
\end{equation}
\begin{equation}
  \mathrm{ELF}(\delta) = 144.75\,\delta^{5} + 71.48\,\delta^{4}
    + 3.14\,\delta^{3} + 0.45\,\delta^{2} + 0.006\,\delta + 0.041
  \label{eq:elf_d}
\end{equation}

The gradient--ELF relationship (Eq.~\ref{eq:elf_g}) is non-monotonic:
ELF rises with increasing slope at moderate curvatures but decreases at
the steepest gradients, reflecting competition between curvature-induced
bond elongation and compressive effects at ripple crests that partially
counteract orbital-overlap reduction.  The shear--ELF relationship
(Eq.~\ref{eq:elf_e}) is symmetric and cubic, consistent with shear
distortions breaking the three-fold bond-angle symmetry of the
$sp^{2}$ lattice and directly misaligning adjacent $p$-orbitals.  Most
significantly, the bond-elongation--ELF relationship
(Eq.~\ref{eq:elf_d}) shows a sharp threshold at $\delta \approx 0.1$
(corresponding to $b_{i} \approx 1.56$\,\AA), above which
$\pi$-electron localisation increases steeply.  Since bond-length
fluctuations in the reconstructed structures reach $\delta \approx 0.2$,
the dynamically rippling graphene sheet undergoes repeated, transient
crossings of this threshold --- driving spatially localised electronic
transitions on the millisecond timescale.

\subsection{Dose-dependent reconstruction accuracy and critical threshold}

We systematically investigate the dependence of reconstruction accuracy on electron dose using simulated datasets spanning a wide dose range. The same graphene model employed in the validation experiments is used throughout to ensure consistency.

\begin{figure}[htbp]
  \centering
  \includegraphics[width=\textwidth]{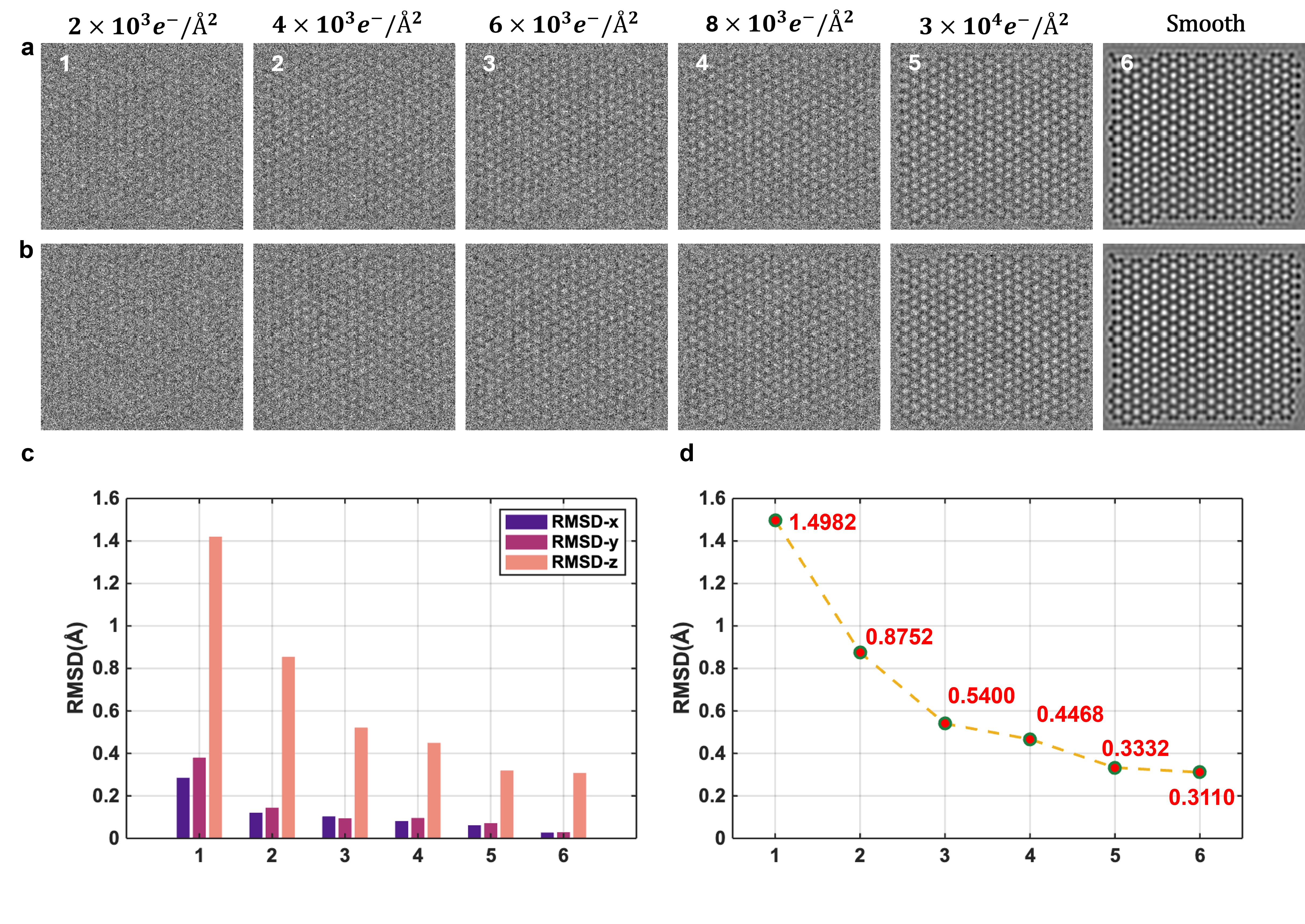}
  \caption{\textbf{Dose-dependent reconstruction accuracy.}
  \textbf{a}, Simulated TEM images at six dose levels: 1,
  $2\times10^{3}$; 2, $4\times10^{3}$; 3, $6\times10^{3}$; 4,
  $8\times10^{3}$; 5, $3\times10^{4}$\,e$^{-}$/\AA$^{2}$;
  6, noise-free.
  \textbf{b}, Reconstructed TEM images from each simulated dataset.
  \textbf{c}, RMSD of $x$, $y$, $z$ atomic coordinates vs.\ dose level.
  \textbf{d}, The Root Mean Squared Error (RMSE) for each reconstructed 3D model.}
  \label{fig:7}
\end{figure}

To identify the lower bound for reliable reconstruction, we first analyse the statistical characteristics of simulated images with doses ranging from $2\times10^{3}$ to $8\times10^{3}$\,e$^{-}$/\text{\AA}$^{2}$. As the dose decreases, the signal-to-noise ratio (SNR) degrades rapidly. In particular, at approximately $4\times10^{3}$\,e$^{-}$/\text{\AA}$^{2}$, the structural signal (manifested as a distinct peak associated with atomic features) becomes comparable to the Gaussian noise distribution and is no longer clearly distinguishable. This behaviour indicates a critical threshold below which structural information is effectively lost. Based on this analysis, we generate simulated TEM datasets at electron doses of $2\times10^{3}$, $4\times10^{3}$, $6\times10^{3}$, $8\times10^{3}$, $3\times10^{4}$\,e$^{-}$/\text{\AA}$^{2}$, and an ideal noise-free limit. The corresponding images (Fig.~\ref{fig:7}a) show progressive degradation of structural contrast with decreasing dose. The reconstructed structures (Fig.~\ref{fig:7}b) are evaluated using the root mean square deviation (RMSD) of atomic coordinates and the root mean squared error (RMSE) of the 3D model (Fig.~\ref{fig:7}c--d).

At high dose ($\geq 3\times10^{4}$\,e$^{-}$/\text{\AA}$^{2}$), the reconstruction error approaches $\sim0.33$\,\AA, with the dominant out-of-plane component reduced to $\sim0.32$\,\AA. In the noise-free limit, the error further decreases to $\sim0.31$\,\AA, representing the intrinsic accuracy limit of the framework. As the dose decreases, reconstruction accuracy deteriorates markedly. At the critical threshold of $4\times10^{3}$\,e$^{-}$/\text{\AA}$^{2}$, the model error increases to $\sim0.87$\,\AA, indicating that the structural signal is insufficient for precise reconstruction. When the dose is reduced to $2\times10^{3}$\,e$^{-}$/\text{\AA}$^{2}$, the error rises to $\sim1.5$\,\AA, approaching the uncertainty of the initial model and indicating that further optimisation becomes ineffective. These results establish $4\times10^{3}$\,e$^{-}$/\text{\AA}$^{2}$ as a practical lower limit for meaningful reconstruction. For high-accuracy results, the electron dose should exceed $6\times10^{3}$\,e$^{-}$/\text{\AA}$^{2}$, while doses near $8\times10^{3}$\,e$^{-}$/\text{\AA}$^{2}$ provide a favourable balance between reconstruction fidelity and minimisation of beam-induced damage.

Overall, this analysis quantifies the trade-off between electron dose, image quality, and reconstruction accuracy, providing practical guidelines for low-dose TEM experiments in beam-sensitive two-dimensional materials.

\section{Discussion}

\textbf{A physics-informed solution to an ill-posed inverse problem.}
Reconstructing 3D atomic coordinates from a single 
two-dimensional (2D) low-dose projection is inherently ill-posed: multiple 
degrees of freedom must be recovered from a single noisy measurement without phase information. 
The proposed framework addresses this challenge through the integration of 
simulated annealing (SA) and molecular dynamics (MD) regularisation. SA enables global optimisation by escaping local minima via the Metropolis acceptance criterion, 
while MD constrains the solution space to physically admissible configurations governed by interatomic potentials. Unlike purely mathematical regularisation, 
this physics-based constraint ensures structural realism under extremely low signal-to-noise conditions. In addition, KL-divergence-based dose calibration aligns the forward model with experimental image statistics, eliminating systematic bias and ensuring 
that the optimisation cost function reflects true structural discrepancy.

\textbf{Structure--property relationships from experimental dynamics.}
The combination of strain tensor analysis, curvature mapping, bond-length characterisation, and density functional theory (DFT)-based electron localisation function (ELF) calculations, all derived from the same reconstructed atomic configurations, constitutes a complete structure--property analysis pipeline applied to a dynamically evolving two-dimensional material. The polynomial relationships established here provide, to our knowledge, the first quantitative mapping between local atomic geometry and electronic structure derived directly from experimental data. In particular, the threshold behaviour in the bond-length dependence of ELF indicates that dynamically rippling graphene undergoes repeated, spatially localised transitions to more strongly localised electronic states. This intrinsic, geometry-driven electronic heterogeneity provides a physically grounded explanation for variability in carrier transport observed in suspended graphene systems.

\textbf{Relationship to prior work.}
Compared with the through-focus pattern-matching approach of Segawa et~al.\cite{RN36}, which addresses the same system, the present framework improves out-of-plane accuracy from $\pm1.0$\,\AA\ to $0.45$\,\AA\ (a $\sim$2.2-fold improvement), while reducing the number of required images from 15 to a single frame and improving temporal resolution from $\sim$70\,s to 1\,ms. In Segawa et~al., specimen drift of $\sim$15--20\,\AA\ over the acquisition window leads to quasi-static reconstructions rather than instantaneous atomic configurations. In contrast, the single-frame formulation used here is intrinsically insensitive to inter-frame drift. The improvement in accuracy arises from the physics-informed optimisation framework, in which simulated annealing iteratively minimises the full pixel-wise $\chi^{2}$ discrepancy against forward-simulated images, with molecular dynamics enforcing physically admissible configurations. By comparison, pattern-matching approaches based on precomputed libraries operate at fixed imaging conditions and do not incorporate iterative refinement under low signal-to-noise conditions. Alternative single-frame methods, such as the STEM-based approach of Li Songge et al.\cite{RN61}, achieve comparable temporal resolution but rely on material-specific contrast mechanisms that are not applicable to monolayer graphene. Bayesian or genetic optimisation strategies\cite{RN62} have been demonstrated for crystalline nanoparticles in Z-contrast STEM, where atom-column information provides additional constraints that are absent in two-dimensional monolayer systems. The present framework is general and applicable to a wide range of two-dimensional materials without requiring system-specific contrast assumptions.

\textbf{Limitations and outlook.}
The current implementation requires several simulated annealing iterations per frame, each involving forward TEM simulation and MD relaxation, which limits throughput for large datasets. Future work may address this by training machine-learning surrogates on the validated reconstructions, enabling faster inference while retaining physical fidelity. Extension to other beam-sensitive two-dimensional materials, such as hexagonal boron nitride and transition metal dichalcogenides, is straightforward, requiring only modification of the interatomic potential in the MD regularisation step. Integration with complementary spectroscopic techniques, such as electron energy-loss spectroscopy, could further enable simultaneous structural, chemical, and electronic characterisation at the atomic scale.

Overall, the framework establishes a general approach for extracting dynamic 3D atomic information from single low-dose TEM images, opening new opportunities for the study of beam-sensitive materials with high temporal and spatial resolution.

\section{Conclusion}

We have demonstrated a physics-informed computational framework that
recovers accurate 3D atomic coordinates of free-standing graphene from
single low-dose TEM images and delivers a complete structure-to-property
analysis from the same data.  The SA+MD inverse solver, anchored by
KL-divergence dose calibration, achieves 0.45\,\AA\ out-of-plane
accuracy at $8\times10^{3}$\,e$^{-}$/\AA$^{2}$ and 1\,ms temporal
resolution.  For the first time, quantitative polynomial relationships
between local surface geometry and electron localisation have been
established from experimentally reconstructed, dynamically evolving
atomic configurations.  A critical dose threshold is identified below
which structural recovery is statistically impossible.  The framework
establishes a general template for atomic-scale structure--property
characterisation of beam-sensitive two-dimensional materials.

\section*{Methods}

\subsection*{Dose calibration by KL divergence}
The experimental electron dose was determined by minimising the KL
divergence $D_{\mathrm{KL}}(P\|Q) = \sum_{i} P(i)\log[P(i)/Q(i)]$
between the normalised pixel-intensity histogram of the experimental
reference frame ($P$) and those of sparse simulated images ($Q$) at
trial doses.  Sparse images model stochastic electron detection by
probabilistic sampling of the normalised simulated intensity distribution
using $N_{e}$ electron events, where $N_{e}$ is set by the trial dose.

\subsection*{Image preprocessing}
Raw frames underwent flat-field correction (Gaussian background
estimation, $\sigma=20$ pixels for $256\times256$ frames; subtracted),
dead-pixel removal ($>5\sigma$ from local mean; replaced by eight-
neighbour mean), BM3D denoising\cite{RN38}, and five-frame
averaging for initial model estimation.

\subsection*{Initial model estimation}
In-plane coordinates were determined by multiple 2D Gaussian fitting.
Out-of-plane coordinates were initialised by the PCD
approximation\cite{RN23} with LOWESS smoothing\cite{RN59}
to suppress outliers.  Ambiguous regions used MAP
estimation\cite{RN63}.  Implemented in MATLAB/StatSTEM\cite{RN64}.

\subsection*{Inverse problem formulation}

Recovering 3D atomic coordinates from a single 2D low-dose TEM image is
a severely ill-posed inverse problem.  Let $\mathbf{r} \in \mathbb{R}^{3N}$
denote the concatenated 3D coordinates of all $N$ atoms in the system.
The forward model $\mathcal{F}: \mathbb{R}^{3N} \to \mathbb{R}^{M}$ maps
an atomic configuration to the $M$-pixel TEM image that it would produce
under known experimental conditions (accelerating voltage, aberration
coefficients, electron dose):
\begin{equation}
  \mathbf{I}^{\mathrm{obs}} = \mathcal{F}(\mathbf{r}) + \boldsymbol{\eta},
  \label{eq:forward}
\end{equation}
where $\boldsymbol{\eta}$ represents shot noise.  The reconstruction task
is to find the configuration $\mathbf{r}^{*}$ that best explains the
observation:
\begin{equation}
  \mathbf{r}^{*} = \arg\min_{\mathbf{r} \in \mathcal{C}}\;
    \mathcal{L}(\mathbf{r}),
  \label{eq:ip}
\end{equation}
where the data-fidelity loss is
\begin{equation}
  \mathcal{L}(\mathbf{r}) = \chi^{2}(\mathbf{r})
    = \frac{1}{M}\sum_{i=1}^{M}
      \bigl(I_{i}^{\mathrm{sim}}(\mathbf{r}) -
            I_{i}^{\mathrm{exp}}\bigr)^{2},
  \label{eq:chi2}
\end{equation}
and $\mathcal{C} \subset \mathbb{R}^{3N}$ is the \emph{admissible set}
of physically plausible atomic configurations.

The problem is ill-posed for three compounding reasons.  First, the map
$\mathcal{F}$ is many-to-one: the projected 2D image integrates along the
beam direction, so all $z$-coordinate information must be recovered from
subtle contrast variations with no direct phase measurement.  Second, at
$8\times10^{3}$\,e$^{-}$/\AA$^{2}$ the image SNR is below~3, meaning
that $\boldsymbol{\eta}$ is of comparable magnitude to the structural
signal.  Third, the landscape of $\mathcal{L}$ contains many local minima
corresponding to structurally distinct but image-indistinguishable
configurations.

Two complementary strategies are required: a global optimisation scheme
that can escape local minima, and a physics-based constraint that defines
$\mathcal{C}$ and prevents convergence to unphysical solutions.  Simulated
annealing provides the former; molecular dynamics regularisation provides
the latter.

\subsection*{Simulated annealing optimisation}

Simulated annealing (SA)\cite{RN37} is a stochastic global optimisation
algorithm that draws its analogy from the physical process of slowly
cooling a solid to its lowest-energy crystalline state.  In the context
of Eq.~\ref{eq:ip}, the configuration $\mathbf{r}$ plays the role of the
system state, and $\mathcal{L}(\mathbf{r})$ plays the role of the energy.

At each iteration $k$, a candidate state
$\mathbf{r}^{*} = \mathbf{r} + \boldsymbol{\xi}$ is generated by
applying a random Gaussian perturbation $\boldsymbol{\xi}$ to the
current atomic positions.  The candidate is accepted according to the
Metropolis criterion:
\begin{equation}
  P(\Delta\mathcal{L},\, T) =
  \begin{cases}
    1 & \Delta\mathcal{L} < 0, \\
    \exp\!\bigl({-\Delta\mathcal{L}}/{T(k)}\bigr) & \Delta\mathcal{L} \geq 0,
  \end{cases}
  \label{eq:metropolis}
\end{equation}
where $\Delta\mathcal{L} = \mathcal{L}(\mathbf{r}^{*}) - \mathcal{L}(\mathbf{r})$
and $T(k) = T_{0}\,\alpha^{k}$ is the temperature following an
exponential cooling schedule with initial temperature $T_{0}$ and
cooling rate $\alpha \in (0,1)$.  At high temperature, the algorithm
accepts uphill moves with non-negligible probability, enabling escape
from shallow local minima.  As $T \to 0$, only downhill moves are
accepted and the algorithm converges to a local minimum near the
current configuration.

The key advantage of SA for this problem is its \emph{global search
capability without gradient information}.  The forward model
$\mathcal{F}$ is evaluated by the Tempas multislice
simulator\,---\,a non-differentiable black-box function of atomic
positions\,---\,making gradient-based methods inapplicable.  SA
requires only function evaluations of $\mathcal{L}$, making it
the natural choice for this setting.  The inner-loop algorithm is
illustrated in Fig.~\ref{fig:8}a.

Termination is triggered when $|\Delta\chi^{2}|$ falls below a
predetermined tolerance over consecutive iterations, indicating that
the configuration can no longer be meaningfully refined.

\subsection*{Molecular dynamics regularisation}

The admissible set $\mathcal{C}$ is defined implicitly by the
condition that atomic positions must be consistent with the known
interatomic potential energy surface of graphene.  We enforce this
constraint by applying molecular dynamics (MD) relaxation after every
SA perturbation step, before the Metropolis acceptance test.

Specifically, after generating the candidate configuration
$\mathbf{r}^{*}$, we perform energy minimisation followed by NVT
equilibration using LAMMPS\cite{RN44} with the Tersoff
potential\cite{RN45,RN46,RN47} (periodic boundary conditions in all
three directions).  The structure is first relaxed to a local energy
minimum on the interatomic potential surface, then equilibrated at
300--1000\,K over 50\,ps using the Nos\'{e}--Hoover
thermostat\cite{RN48,RN49} (1\,fs timestep).  Final coordinates are
obtained by averaging over the NVT trajectory sampled every 0.1\,ps.
The resulting coordinates $\tilde{\mathbf{r}}^{*}$ replace
$\mathbf{r}^{*}$ in the Metropolis criterion.

Critically, MD regularisation acts as a \emph{projection} of the SA
candidate onto $\mathcal{C}$: regardless of how improbable the
perturbed geometry is, the relaxed structure always satisfies the
constraints imposed by the interatomic potential.  This is
fundamentally different from soft mathematical regularisers (e.g.\
Tikhonov regularisation or total variation), which penalise unlikely
configurations but cannot guarantee physical admissibility.  Under
the extremely low SNR conditions of this experiment, where noise can
drive unconstrained optimisation far into unphysical regions of
coordinate space, this hard-constraint property is essential.

\begin{figure}[htbp]
  \centering
  \begin{minipage}[t]{0.47\textwidth}
    \centering
    \textbf{a}\\[6pt]
    \includegraphics[width=\textwidth]{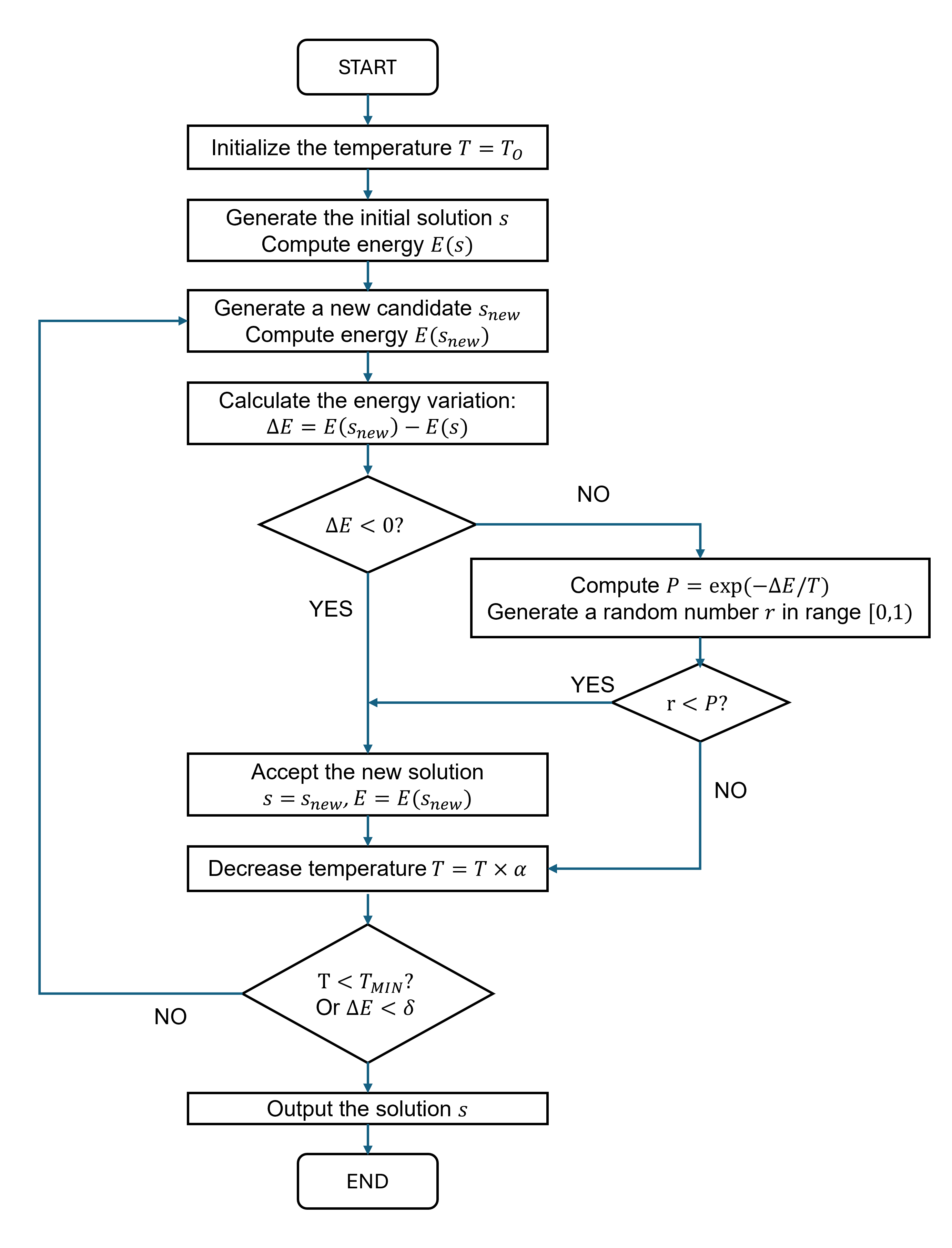}
  \end{minipage}
  \hfill
  \begin{minipage}[t]{0.47\textwidth}
    \centering
    \textbf{b}\\[6pt]
    \includegraphics[width=\textwidth]{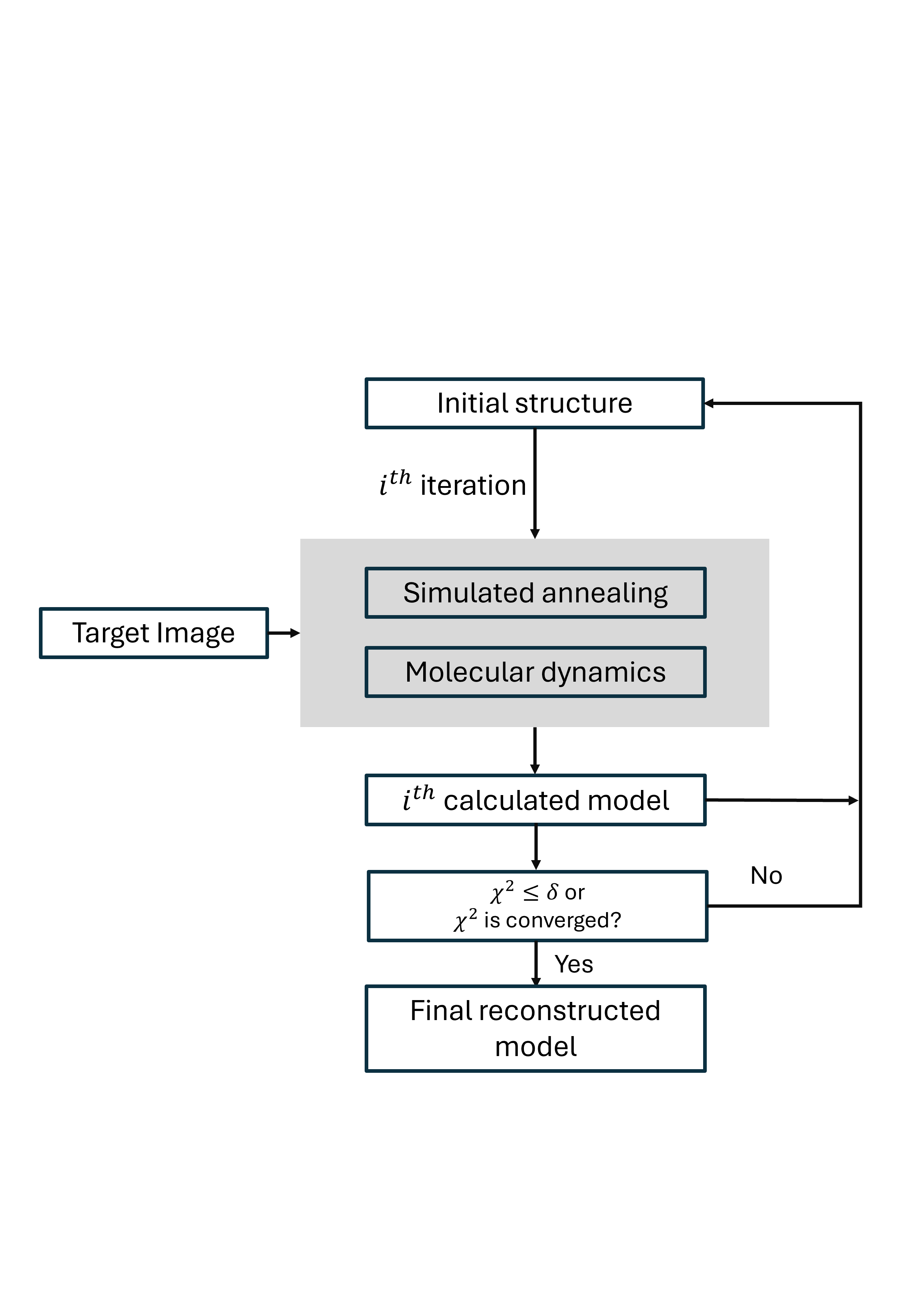}
  \end{minipage}
  \caption{\textbf{Algorithm flowcharts for the SA+MD reconstruction
  framework.}
  \textbf{a}, Inner-loop simulated annealing (SA) algorithm.
  Starting from an initial temperature $T_{0}$ and atomic configuration
  $s$, a candidate state $s_{\mathrm{new}}$ is generated at each
  iteration and evaluated via the forward TEM simulator.  If the energy
  (image discrepancy $\chi^{2}$) decreases ($\Delta E < 0$), the move
  is accepted unconditionally; otherwise it is accepted with Boltzmann
  probability $P = \exp(-\Delta E / T)$, allowing escape from local
  minima at high temperature.  The temperature decreases geometrically
  ($T \leftarrow T \times \alpha$) until the termination criterion
  $T < T_{\min}$ or $|\Delta E| < \delta$ is satisfied.
  \textbf{b}, Outer-loop SA+MD iteration.  At each outer iteration $i$,
  the current atomic model is passed to the SA+MD block (shaded): SA
  proposes a coordinate perturbation and MD relaxation enforces physical
  admissibility by projecting the candidate onto the interatomic
  potential energy surface.  The resulting $i$-th calculated model is
  evaluated against the target TEM image; if $\chi^{2} \leq \delta$
  or $\chi^{2}$ has converged, the loop terminates and the final
  reconstructed model is returned; otherwise the model is passed back
  as the initial structure for the next iteration.}
  \label{fig:8}
\end{figure}

The combined SA+MD outer loop is illustrated in Fig.~\ref{fig:8}b.
Each outer iteration corresponds to one complete SA+MD cycle applied
to the full atomic configuration.  Convergence is assessed at the
outer-loop level by monitoring $\chi^{2}$ between successive cycles;
the reconstruction terminates when $\chi^{2}$ change falls below
threshold or a maximum number of outer iterations is reached.

\subsection*{Strain analysis}
Atomic displacements were obtained by nearest-neighbour matching to a
perfect flat graphene reference.  Strain components
$\epsilon_{xx} = \partial u/\partial x$,
$\epsilon_{yy} = \partial v/\partial y$,
$\epsilon_{xy} = \frac{1}{2}(\partial u/\partial y +
\partial v/\partial x)$
were computed by finite differences on the interpolated displacement
field.  For flattened models, global tilt was removed by least-squares
plane fitting and rotation to $z=0$ before displacement computation.

\subsection*{Gradient and bond-length analysis}
The $z$-height surface $F(x,y)$ was obtained by bicubic interpolation
of 3D atomic positions; gradient components were evaluated by finite
differences.  Bond-length change $\delta = (b_{i}-b_{0})/b_{0}$ was
computed at each C--C midpoint and a 3D surface interpolated through
$(x_{b}, y_{b}, \delta)$ for spatial visualisation.

\subsection*{DFT and ELF calculations}
Calculations used VASP\cite{RN50,RN51} with the PBE
functional\cite{RN52} and PAW
pseudopotentials\cite{RN53,RN54}.  Cell:
$a=b=60.85$\,\AA, $c=25.00$\,\AA; atoms fixed at reconstructed
positions.  $\Gamma$-point sampling (Monkhorst--Pack\cite{RN55});
Gaussian smearing 0.05\,eV; energy convergence $10^{-5}$\,eV;
plane-wave cutoff 450\,eV.  The ELF\cite{RN56,RN57} was
visualised with VESTA\cite{RN58}.

\section*{Data availability}
The experimental TEM data that support the findings of this study are
available from the corresponding authors upon reasonable request.
Simulated datasets and all analysis outputs are available in the
project repository listed under Code availability.

\section*{Code availability}
The full reconstruction framework, including preprocessing scripts,
SA optimisation, strain/gradient\\
/bond-length analysis utilities, and
DFT post-processing tools, is openly available at
\url{https://github.com/xjzhang2365/3D-Reconstruction-Low-Dose-Imaging}.

\section*{Acknowledgements}
Funding information will be updated prior to journal submission.

\section*{Author contributions}
X.Z. conceived the inverse-problem framework, developed and implemented all algorithms, performed the reconstructions and analyses, and wrote the manuscript. A.I.K. acquired the experimental TEM data and contributed to scientific discussion. F.-R.C. facilitated access to the experimental data and supervised the research. S.-W.H. contributed to algorithm development and validation. Y.W. and J.-P.C. performed the DFT and ELF calculations. R.K. provided the Tempas simulation environment.

\section*{Competing interests}
R.K.\ is the developer of the Tempas software used for TEM image simulation and forward modelling in this work, and is the owner of Total Resolution LLC, from which a software licence was purchased by the laboratory of F.-R.C.\ at City University of Hong Kong. This relationship constitutes a potential competing financial interest. All other authors declare no competing interests.

\bibliographystyle{unsrt}
\bibliography{references}

\end{document}